\documentclass{ws-procs9x6}
\newcommand\beq{\begin{equation}}
\newcommand\eeq{\end{equation}}

\begin{document}

\title{NONPERTURBATIVE PHYSICS AT SHORT DISTANCES}

\author{V.\,I.\,ZAKHAROV}

\address{ Istituto Nazionale di Fisica Nucleare - Sezione di Pisa,
Largo Pontecorvo, 3, 56127 Pisa, Italy;\\
Max-Planck Institut f\"ur Physik, F\"ohringer Ring 6, 80805, M\"unchen,
Germany }

\begin{abstract}
There is accumulating evidence in lattice QCD that 
attempts to locate confining fields in  vacuum configurations
bring results explicitly depending on the lattice spacing
(that is, ultraviolet cut off).
Generically, one deals with low-dimensional vacuum defects
which occupy a vanishing fraction of the total four-dimensional space.
We review briefly existing data on the vacuum defects and
their significance for confinement and other non-perturbative
phenomena. We introduce the notion  of `quantum numbers'
of the defects and draw an analogy, rather formal one, to
developments  which took place about 50 years ago and were
triggered by creation of the Sakata model.
\end{abstract}


\bodymatter
%\maketitle

%\thispagestyle{fancy}
%\setcounter{page}{0}

\section{Introduction}

This mini-review was prepared for the Conference
celebrating 50 years of the Sakata model \cite{sakata}
held at the University of  Nagoya in November 2006. 
This celebration  is an emotional event for me since it 
happened so that I learnt about the Sakata model before
I had exams in field theory.
In the winter 1960-1961
  L.B. Okun was giving at the Institute of Theoretical and Experimental Physics (ITEP)
a course of
lectures on the Sakata model and weak interactions.
By chance,  I attended the  lectures. Having 
 no education in field theory did not help much 
 to understand physics but I do remember well
 the excitement and tense expectations of imminent
 and decisive developments.
 
 Later, I had ample opportunity to learn more about the model
 since there was exploding activity at ITEP in that direction.
In particular, my theses adviser, I.Yu. Kobzarev
was a close collaborator of L.B. Okun. There is no need, however,
 to go into historical remarks here 
 \footnote{It is difficult for me to leave the recollections grounds without 
 mentioning the role, vision and devotion to physics of I.Ya Pomeranchuk
 which are beginning and end of all my memories of that time.}
 since there appeared
 recently 
 recollections of L.B. Okun \cite{okun}, to be published in the same Proceedings.

Instead, we will proceed directly to the topic of the talk,
that is non-perturbative physics at short distances.
The connection to the theme of the Conference, as we see it, is
as follows. The Sakata model  was a pre-runner of the quark
model. And quarks acquired solid theoretical status only when
it was realized that they survive  at short distances.  
During last five years or so, evidence has been accumulating 
that confining fields in QCD are actually hard, i.e. survive at short
distances.
 Hence, there is a chance that non-perturbative fields could be
 understood in terms of fundamental  physics.
 The analogy to fundamental quarks might look too far fetched
  or even artificial.
The reason to draw parallels to sakatons or quarks
is to convey amusement brought by the lattice data
and facilitate discussion of the data by the continuum community.

The Sakata model  explored the idea
that there exist fundamental constituents of matter.
Roughly speaking, the messages were:
\begin{itemize}
\item{ There might exist a few fundamental constituents.
Earlier, one rather thought  in terms of a single fundamental field
(if at all)}
\item{Even in the absence of detailed dynamical theory one could check
 the choice of the fundamentals  by selection rules}
\end{itemize}
Later, there was found a simple and still meaningful limit to
consider dynamics. Namely:
\begin{itemize}
\item{ Constituents fill in an
irreducible representation of a simple group, let it be associated with a broken symmetry.
}
\end{itemize}

Trying to build up parallels for the non-perturbative physics
we will proceed in the following way:
\begin{itemize}
\item{review briefly the evidence in favor of vacuum defects
in Yang-Mills theories (mostly, $SU(2)$ and $SU(3)$ cases)
and try to identify basic ones.}
\item{Introduce notion of quantum numbers for non-perturbative  
 structures, like dimensionality of the vacuum defects}
 \item{Speculate that the quantum numbers of the defects can be predicted by
 studying classical solutions of large-$N_{c}$  theories.}
 \end{itemize}
 
 \section{Phenomenology of vacuum defects}
 \subsection{Critical exponents}
 
 In a pure Yang-Mills theory, everything can be  expressed in terms of the running
 coupling constant.  Imagine that measurements are performed with 
 resolution $a$. In lattice formulation, an example of such a measurement is a Monte-Carlo
 generated set of vacuum gauge fields $\{A_{\mu}^{a}(x)\}$.
 The lattice spacing $a$ plays the role of resolution of measurements
 on the  vacuum fields.
 
 The coupling normalized at the lattice spacing,
$g^{2}(a)$ is then the relevant  parameter. Because of asymptotic
 freedom $\lim_{a\to 0} g^{2}(a) \to 0~$,
 and the vacuum fields 
 $\{A_{\mu}^{a}(x)\}$ are, to zero approximation, zero-point fluctuations
 of the gluonic fields.
 One can include also perturbative corrections, order by order.
 The probability of a non-perturbative, or barrier transition
 is generically of order
 \begin{equation}\label{expectation}
 \theta_{non-pert}~\sim~\exp(-const/g^{2}(a))~\sim~(a\cdot \Lambda_{QCD})^{\alpha}~~,
  \end{equation}
 where $\alpha >0$ and can be called a critical exponent.
 
 The terminology is borrowed from theory
 of percolation (see, e.g., \cite{grimmelt}) and with a good reason.  
Theory of percolation refers to second order phase transitions.
 The most interesting case is the so called supercritical phase when 
 there exists a vacuum condensate, or a tachyonic mode. If the phase transition
 corresponds, in some parameterization to $\epsilon=0$ then for small $\epsilon$
 many observables are described in terms
 of critical exponents. For example, density of condensate
 is proportional to
 \begin{equation}\label{theta}
 \theta_{condensate}~\sim~\epsilon^{\alpha}~~,~~\alpha~>~0~.
 \end{equation}
In  percolation theory, or in the  lattice language Eq. (\ref{theta})
refers to the probability of a given link to belong to an infinite cluster.
The infinite cluster is a substitution for vacuum condensates of the continuum
theory.

Moreover, as is well known,
the continuum limit, or $a~\to~0$  corresponds to the point of second
 order phase transition. Indeed, the correlation length in  lattice units
 should tend to infinity with $a~\to~0$. This is the condition that 
at physical scale there is no dependence on the lattice spacing.
In dimensionless units, the product $(a\cdot \Lambda_{QCD})$
quantifies closeness to the point of the phase transition.

Next, why the Yang-Mills theories correspond to 
supercritical phase of the percolation theory.
The point is that confinement is due to instability of the perturbative
vacuum. The instability is condensation of magnetic degrees of freedom.
Thus,  we come to expectations
(\ref{expectation}).

Note that 
smallness of the probability  (\ref{expectation}), implies singular
non-perturbative fields in same measurements. Indeed, imagine that
there is a conserved quantity, expressed in terms of non-Abelian fields.
Then constancy of the product of probability (\ref{expectation}) and of the field strength
implies singular fields. We will have examples later.
 
 \subsection{Lattice data}
 
 In this section we will enumerate various critical exponents,
concentrating on definitions and without trying to 
immediately interpret the data.
\begin{itemize}
\item{Magnetic strings} 
\end{itemize}
The strings are defined as closed, infinitely thin surfaces
in the vacuum state which can be open on a 't Hooft line.
The 't Hooft line, in turn, is nothing else but trajectory of an external 
magnetic monopole. Lattice definition of the external  monopole
is the end point of a Dirac string which carries quantized magnetic flux,
determined in terms of the center group.

As is argued in detail in \cite{vz}, magnetic strings defined
in this way could be identified with the so called center
vortices, for review see \cite{greensite}.

The surfaces percolate through the vacuum. For the probability of a given plaquette
to belong to the surfaces one finds: \cite{greensite,ft}
\begin{equation}\label{first}
\theta_{{plaquette}}~\sim~(a\cdot\Lambda_{QCD})^{\alpha_{1}}~~,~~
\alpha_{1}~\approx~2~~.\end{equation}
Alternatively, Eq (\ref{first}) can be expressed as  statement that
the total area of the surfaces is in physical units:
\begin{equation}\label{area}
(Area)_{strings}~\sim~\Lambda_{QCD}^{2}V_{4}~~,
\end{equation}
where $V_{4}$ is the total volume of the lattice.
 
 One can measure also extra action associated with the strings \cite{ft}:
 \begin{equation}\label{singular}
 S_{strings}\sim~ (\Lambda_{QCD}^4V_4)(a\cdot\Lambda_{{QCD}})^{-\beta_{1}}
\sim~\Lambda_{QCD}^2a^{-2}V_{tot},~~ \beta_{1}~\approx~2~.
 \end{equation}
  Thus, the non-perturbative fields associated with the strings 
  are singular in the continuum limit of $a\to 0$
(as promised above).
  
  To appreciate, which conserved quantity is hiding behind the observation $\alpha_1=\beta_1$
(compare (\ref{first}) and (\ref{singular})) let us note that, upon fixing $\alpha_1$
one can derive $\beta_1=\alpha_1$ from the requirement that the string tension determined
from the stochastic model for the Wilson line is of order $\sigma\sim\Lambda_{QCD}^2$
  \cite{vz1}. 
  
 \begin{itemize}
 \item{Monopole clusters}\end{itemize}
 Monopole trajectories, or 1d defects are defined in specific lattice 
 terms, for review see \cite{suzuki}. Namely, one replaces
 the original non-Abelian fields by the closest  Abelian configuration,
 $
 \{A_{\mu}^{a}(x)\}~~\to~~\{A_{\mu}^{3}(x)\}~~,$
    where $a=3$ is the color index (and we consider SU(2) case). 
    The monopoles are defined then as Dirac monopoles in terms of
    the projected fields $\{A_{\mu}^{3}(x)\}$.
    The physical idea is that confinement is
    encoded in Abelian degrees of freedom.
    
    The clusters of monopole trajectories possess remarkable scaling properties.
    In particular, there exists an infinite, or percolating cluster and
    the probability of a given link to belong
    to this cluster is of order  \cite{boyko}:
     \begin{equation}
     \theta_{link}^{perc}~~\sim~~(a\cdot\Lambda_{QCD})^{\alpha_{2}}~~,~~
     \alpha_{2}~\approx~3~.     \end{equation}
     Similar probability in case of finite clusters  is of order:
 \begin{equation}
 \theta_{link}^{finite}~\sim~(a\cdot \Lambda_{QCD})^{\alpha_{3}}~~,~~
 \alpha_{3}~\approx~2~.
 \end{equation}
 
 Monopoles are associated with singular non-Abelian fields \cite{bornyakov}:
 \begin{equation}
 S_{{monopoles}}~~\sim~~(L_{mon}^{tot}\cdot \Lambda_{QCD})(a\cdot\Lambda_{QCD})^{-\beta_{2}}~~,~~\beta_{2}~\approx~1~.
 \end{equation}
 Monopoles give contribution to the total action similar to that of strings.
This is in fact not accidental since monopoles are closely associated with
the surfaces, as we will discuss in more detail later.
 
 \begin{itemize}
 \item{Holographic 3d defects}
  \end{itemize}
 
Consider 3d volumes spanned on 
2d surfaces, or magnetic strings introduced above. 
In 4d this volume is not uniquely defined and one
concentrates on the minimal volume\cite{3d}.
The probability of a given elementary cube on the lattice to belong to the
volume considered turns to be:
\begin{equation}
\theta_{cube}~~\sim~(a\cdot\Lambda_{QCD})^{\alpha_{4}}~~,~~
\alpha_{4}~\approx~1~.
\end{equation}
 This result is important by itself since it 
 means that the surfaces are {\it not} the  so called branched polymers
 (for discussion and references see \cite{vz}). 
 
 The 3d defects are closely related to both
 confinement and spontaneous breaking of chiral symmetry.
 Existence of such a relation is revealed by a remarkable procedure
 introduced in Ref. \cite{delia} \footnote{We describe the procedure in version
 of Ref. \cite{3d}.} Namely, one strongly modifies links
 belonging to the 3d volume considered in an algorithmically
  well defined way:
  \begin{equation}\label{replacement}
  U^{3d}_{\mu}(x)~\to~Z_{\mu}(x)U^{3d}_{\mu}(x)~~,
  \end{equation}
  where no summation over $\mu$ is implied,
  $U_{\mu}(x)$ are original matrices generated with Yang-Mills
  action, and $Z_{\mu}$ are numbers which in SU(2) case are $\pm 1$.
   Whether $Z_{\mu}(x)$ is (+1) or (-1) is determined for
  each particular link in terms of projected fields \cite{delia}.
  
  It would take us too far to explain details of determining $Z_{\mu}$.
  What is important now is that replacement (\ref{replacement})
  is an ad hoc strong modification of the original gauge fields
  applied, however, only to a vanishing (in the limit $a\to 0$)
  fraction of the total volume. 
  
  The result of the substitution
  (\ref{replacement}) is that both confinement and
  quark condensate, $\langle\bar{q}q\rangle$ disappear \cite{delia}.
  This happens only if the 3d volume is chosen in the way described above
(or any gauge transform of it).
    
  Thus, the whole information on the confinement and
  chiral symmetry breaking can be encoded in a 3d volume, which is a vanishing 
  fraction of the total 4d volume. Thus, one can talk about a 'holographic' 3d
  volume.
  
  \begin{itemize}
  \item{Topological defects}
  \end{itemize}
As is well known topology of background gluon fields
is imprinted on fermionic low-lying modes.
  Recently, measurements \cite{random} of the volume occupied  
  by topological fermionic modes attracted a lot
  of attention, for review see \cite{reviews}.
  The measurements found non-trivial dependence on
  the lattice spacing while the expectations-- based on the
  instanton model-- were that the topological modes 
  occupy simply the volume of order $\Lambda^{-4}_{QCD}$.
  
  In more detail, one considers solutions of the eigenvalue
problem
\begin{equation}\label{lambda}
D_{{\mu}}\gamma_{\mu} \psi_\lambda \, =\, \lambda \psi_\lambda\;,
\end{equation}
where the covariant derivative  is constructed on the vacuum
gluonic fields  $\{A_{\mu}^a(x)\}$. 
For exact zero modes,
there is well known index theorem:
\begin{equation}\label{zeromodes}
n_{+}-n_{-}~=~Q_{top}\;.
\end{equation}
Assuming that the topological charge $Q_{top}$ fluctuates independently 
by order unit
on pieces of  4d volumes measured in physical units:
\beq \label{topologicalcharge}
\langle Q_{top}^{2}\rangle~\sim ~\Lambda_{QCD}^{4}V_{tot}.
\end{equation}
The so called near-zero modes occupy, roughly speaking
the interval
\begin{equation}\label{band}
0~~<~~\lambda~~<~~\pi/ L_{latt}~~,
\end{equation}
where $L_{latt}$ is the linear size of the lattice.
These modes determine the value of the quark condensate via
the Banks-Casher relation:
\begin{equation}\label{bankscasher}
\langle \bar{q}q \rangle ~=~ -\pi \rho(\lambda \to 0)\;,
\end{equation}
where  $\lambda \to 0$ with the total volume tending to infinity.

Measurements 
\cite{random} with high resolution, that is  on the original fields
$\{A_{\mu}^{a}(x)\}$ 
confirmed the general relations (\ref{topologicalcharge})
and (\ref{bankscasher}). They also demonstrated
that the volume occupied
by low-lying modes tends to zero in the continuum limit of
vanishing lattice spacing, $a\to 0$:
\begin{equation}\label{shrinking}
\lim_{a\to 0}{V_{mode}}~\sim~(a\cdot \Lambda_{QCD})^{\alpha_{5}}~\to ~0\;,
\end{equation}
where numerically $\alpha_{5}$ is between 1 and 2. 
The volume $V_{mode}$ is defined in terms of the Inverse Participation
Ratio (IPR) \footnote{Independent evidence in favor of shrinking
of the regions occupied  by topologically non-trivial gluon fields 
to 3d volumes was
obtained in \cite{gubarev}.}.

 \begin{itemize}
 \item{Stochastic defects}
 \end{itemize}
    
   Localization properties were also studied for test
   color scalar particles in Yang-Mills vacuum \cite{scalars}.
   Namely, one studies solutions of the equation
   \begin{equation}\label{scalars}
   D^{2}\phi_{\lambda}~=~\lambda^{2}\phi_{\lambda}  ~~.
   \end{equation}
   The  analysis 
   has been performed for SU(2) case  for  color spin
   T=1/2,1,3/2.
   
   In case of scalars, already in the continuum theory one 
   expects a drastic dependence on the lattice spacing due to the  ultraviolet divergent
  radiative mass correction, $
   \delta M^{2}~=~\lambda_{min}^2~\sim~\alpha_sa^{-2}.$
   To describe particles with physical masses, one uses subtractions.
   In particular, the
    subtraction constant $M^{2}=-\lambda_{min}^{2}$ 
  corresponds to zero renormalized mass.
   After fixing the renormalized mass, the theory is fully determined
   perturbatively.
   
   Lattice treatment  \cite{scalars}
   uncovers new phenomena.
   There appear localized states
   which do not propagate at all.
   These states correspond to eigenvalues
   in the interval
   \begin{equation}\label{interval}
   \lambda^{2}_{min}~<~\lambda^{2}~<
   \lambda^{2}_{mob}~~,
   \end{equation}
  where $\lambda^{2}_{mob}$
   is the so called mobility edge.
Probably, the mechanism of the localization in the Yang-Mills case
is similar to the Anderson localization, namely, 
presence
of random, or stochastic fields in the vacuum.

      If we identify the mobility edge with the radiative mass
   and introduce subtraction  $M^{2}=-\lambda_{mob}^{2}$,
     then there is an advantage that higher eigenvalues $\lambda_{n}^{2}>\lambda_{mob}^{2}$ might be associated with
     standard plane waves. 
   However, the
   renormalized eigenvalue 
      $$\tilde{\lambda }^{2}~\equiv~\lambda^{2}~-~M^{2}$$
   is then negative, or tachyonic in the interval (\ref{interval}).
   
   More generally, existence of the two  invariants, $\lambda^{2}_{mob}$
   and $\lambda^{2}_{min}$ associated with  a single particle defies the standard 
   classification scheme of states with respect to the Poincare group.

   For the  localized  states, one  introduces 
   new critical exponents \cite{scalars}:
   \begin{eqnarray}
   \lambda^{2}_{mob}~-~\lambda^{2}_{min}~\sim~\Lambda^{2}_{{QCD}}
   (a\cdot \Lambda_{QCD})^{-\beta (T)}~~~,\\ \nonumber~~~
   V_{loc}(\lambda_{{min}})~\sim~\Lambda_{QCD}^{-4}(a\cdot\Lambda_{QCD})^{\alpha(T)}
   ~,
   \end{eqnarray}
   where $V_{loc}(\lambda_{min})$ is the localization volume.
   
   The exponents $\alpha(T),\beta(T)$ were measured on the lattice
   and the measurements brought once again absolutely unexpected
   results. Namely, the values of $\alpha(T),\beta(T)$ depend crucially on the
   color spin:
   \begin{eqnarray}\label{results}
   \alpha(T=1/2)~~\approx~~0~,~~~\beta(T=1/2)~~\approx~~0~;\\
    \alpha(T=1)~~\approx~~2~,~~~\beta(T=1)~~\approx~~1~\nonumber.
    \end{eqnarray}
   Note that the interval (\ref{interval}) appears to be divergent
   in the continuum limit for the adjoint case. 
   The renormalization program
   for the scalar particles in the adjoint representation   cannot actually
   be performed then \cite{scalars}.
 
\begin{itemize}
\item{Shrinking confining string}
\end{itemize}

Most recently, there appeared evidence 
\cite{bgm} that the confining string, connecting heavy quarks 
shrinks to a mathematical line in the continuum limit:
\begin{equation}
\delta^2~\sim~\Lambda_{QCD}^{-2}(a\cdot \Lambda_{QCD})^{\alpha_6},~~\alpha_6~\approx~2~,
\end{equation}
where $\delta$ is the thickness of the flux tube. 
This is the most challenging observation (compare with the now-standard picture of Ref.\cite{muenster})..

 \section{Glimpses of theory}
 
 \subsection{Dependence on the resolution of measurements}

An element of theory has already been introduced in fact, in the form
of language used. The point is that in the text-book language the lattice
spacing is a substitute for an ultraviolet cut off. We are labelling the
lattice spacing as resolution of measurements, reserving the use
of 'ultraviolet cut off' for the  
divergences in field theory. There are many observables which  can well depend on the resolution
of measurements \cite{arkady}.

Some matrix elements should not depend on the resolution.
In terms of critical exponents they correspond to a vanishing index.
Consider, for example, Banks-Casher  relation (\ref{bankscasher}).
It relates the quark condensate to the density of eigenstates $\rho(\lambda)$:
\begin{equation}\label{bc}
\langle\bar{q}q\rangle~=~\lim_{m\to 0}m\cdot{\int{\rho(\lambda)d\lambda\over( m^2+\lambda^2)}}.
\end{equation}
 With $a\to 0$ the number of modes grows, $\lambda_{max}\sim1/a$. However, all modes
with $\lambda\sim1/a$ drop off from (\ref{bc}) and the only relevant
quantity is $\rho(\lambda\to 0)$. Thus, for the quark condensate
to be independent on $a$ we need,
\begin{equation}
\rho(\lambda\to 0)~\sim~(a\cdot \lambda_{QCD})^{\gamma_1}~,~~\gamma_1~=~0~.
\end{equation}
which is obeyed by the data \cite{random}. 
Another example was mentioned above. Namely, by measuring the Wilson line
one determines the heavy quark potential, $V_{\bar{Q}Q}(r)$. At large
R the potential grows linearly,
$$V_{\bar{Q}Q}(r)~\sim~\sigma \cdot r~,~\sigma~\sim~\Lambda_{QCD}^2
(a\cdot\Lambda_{QCD})^{\gamma_2}~,~\gamma_2~=~0~,$$
and the tension $\sigma$ (as being directly related to free energy)
cannot depend on the lattice spacing.
This condition constrains the critical exponents $\alpha_1,\beta_1$
introduced above, $\alpha_1=\beta_1$.

Note that the Banks -Casher relation does not constrain the volume occupied
by the eigenmnodes with $\lambda\to 0$. Moreover,
there is no other matrix element which would require $a$-independence of
the volume occupied by the topological fermion modes \cite{arkady}.

\subsection{Shrinking topological fermionic modes}

In the preceding subsection we emphasized that - as was realized only recently -
many observables {\it could} depend on the resolution. 
In case of topological fermionic modes it seems possible to make the next step
and argue that the topological modes {\it should} shrink to a vanishing
4d volume \cite{arkady}.

The essence of the argumentation is that instantons represent a non-unitary
contribution to correlator of topological densities \footnote{Analysis of 
the unitarity constraints was given
much earlier, see  \cite{seiler} and references therein. It is only discussion of implication 
for the $a$-dependence of measurements that is recent.}.
For instanton:
\begin{equation}
\langle~ G\tilde{G}(x),~G\tilde{G}(0)~\rangle~>_{instanton~}~>~0,~~x~<~\rho_{inst}~~,
\end{equation} 
where $\rho_{inst}$ is the instanton size.
On the other hand, from unitarity alone one can establish \cite{seiler}
\begin{equation}\label{general}
\langle~ G\tilde{G}(x),~G\tilde{G}(0)~\rangle~>_{unitary}~<~0~.
\end{equation}
Eq. (\ref{general}) means that at any given small $x$ the perturbative contribution
is larger than that of non-trivial topology fluctuations, like instantons.

On the other hand, perturbatively the r.h.s. of Eq (\ref{topologicalcharge})
is vanishing. The only way to reconcile these, apparently conflicting 
 constraints is to assume for the non-perturbative contribution
a singular form: 
\begin{equation}\label{delta}
\langle~ G\tilde{G}(x),~G\tilde{G}(0)~\rangle_{non-pert}~\sim~\delta^4(x)\Lambda_{{QCD}}^{4}~,
\end{equation} 
 for checks on the lattice see \cite{digiacomo}.

In the language of dispersion relations Eq. (\ref{delta})
implies that the non-perturbative effects are encoded in a subtraction term \cite{seiler}.
In terms of topological fermionic modes Eq (\ref{delta})
implies that the modes are shrinking to a `subtraction volume' which is 
a vanishing part of the total 4d volume \cite{arkady}. With a stretch of imagination one can argue that
they shrink, most probably, to a 3d volume \cite{arkady}.

\section{Quantum numbers of vacuum defects}

\subsection{Quantum numbers of 2d defects}

The argumentation outlined briefly in the preceding section
demonstrates that low-dimensional vacuum defects 
are in no way a lattice curiosity but could have been predicted within
field theory. So far, however, the field-theoretic argumentation
was found only in case of topological defects. Thus, we still have to mostly rely
on the phenomenology to consider the vacuum defects. We will label the defects with
quantum numbers. In this subsection we
will consider 2d defects, or lattice strings.

{\it Dimensionality}

The first, and obvious candidate for a quantum number is a dimensionality
itself. In particular, for strings it is $d=2$. Note that we do not admit
an anomalous fractional dimension of defects. In particular, the surfaces has
dimension 2 in physical units, see Eq. (\ref{area}).

{\it Abelianity}

Non-Abelian fields living on a 2d surface are in  fact Abelian. This is not an approximation
but a general observation \footnote{In all the generality, the observation is made
and explored in \cite{gukov}. Earlier, it was explored in context of more concrete
applications, see in particular \cite{gubarev1}.
Let is also note that the surfaces considered in \cite{gukov}
can be open on monopole lines, with unquantized monopole charges.
The surfaces relevant to confinement can be open on the monopole lines
with quantized charges, although their own flux is not quantized.}.
 
 Indeed, consider a plane with coordinates $x_3=x_4=0$.
Then, by non-Abelian field living on this surface one would understand
$G_{12}^a~\neq~0~$,
where $a$ is the color index. Using gauge invariance one always can rotate
this field to a given direction in the color space,
$$G_{12}^a~\to~G_{12}^3~~,$$
where, for simplicity of notations we consider SU(2) gauge group.

{\it Chirality}

Thus, the Abelian nature of fields living on a 2d surface looks obvious.
There is a subtle further point, however, and it concerns chirality of the
gluon fields. The point is that in the 4d Euclidean space the rotation (Lorentz) group 
is a direct product 
$$O(4)~=~O(3)\times O(3)~~,$$
and it is only natural to consider irreducible representation with respect to these groups, which are $G^{3}_{12}\pm G^{3}_{34}$. Reserving for an arbitrary combination of the two, we come to the conclusion that 2d surfaces are characterized
by two invariants which can be taken $G^{2}$ and $G\tilde{G}$,
none of which is being quantized. 
In other words, gluon fields living on the surfaces are chiral, generally
speaking. It goes without saying that if the surfaces percolate
thorough the vacuum, their chirality varies.

\subsection{Two-dimensional defects as fundamental ones}

Let us check whether all the data on the critical exponents,
which we listed above,
 can be understood
in terms of 2d surfaces. 
{\it Monopole clusters} were defined through Abelian projection. 
At first sight, the Abelian projection has nothing to do with the surfaces.
However, now we have learned that surfaces are endowed in fact
with Abelian fields. Thus, one could speculate that the surfaces should be seen
in the Abelian projection as well (since we should not loose the surfaces by
abelianization).

These expectation turn to be true: {\it within error bars, monopole trajectories lie
 on the vortices}, see \cite{ft} and references therein.
 Thus monopoles could be manifestations of the same surfaces and just tell us that
 the surfaces are structured with trajectories.
 
 In the schemes with 2d surfaces as fundamental defects, the
 {\it holografic 3d defects} become excessive. Their role is reduced to the role 
 played by their boundaries. 
 
 The next question is whether the 
 surfaces carry chirality and explain  {\it topological defects}. In the context of the
 lattice measurements, the question is whether the topological fermionic modes,
 with improving resolution
 shrink to the surfaces.
 
 An attempt to answer this question was undertaken in
Ref. \cite{correlator}
through a direct  study of correlation between intensities $\rho_{\lambda}(x)$ of
fermionic modes with eigenvalue $\lambda$ and of vortices.
In more detail,
center vortex is a set of plaquettes $\{D_i\}$ on the dual lattice. Let us
denote a set of plaquettes dual to $\{D_i\}$ by $\{P_i\}$. Then the
correlator in point is defined as:
\begin{eqnarray}\label{correlator}
C_\lambda(P) =
\frac{\sum_{P_i} \sum_{x \in P_i} (\rho_\lambda(x) -\langle \rho_\lambda(x) \rangle ) }
{\sum_{P_i} \sum_{x \in P_i} \langle \rho_\lambda(x) \rangle} \, .
\label{eq:z2_plaq_corr_orig}
\end{eqnarray}
The data does demonstrate  strong positive correlation
between intensities of topological modes and
density of vortices nearby.
Moreover,  the value of the correlator
depends on the eigenvalue
and the correlation is strong only for the topological fermionic modes.
Although the data does show that the correlator grows for smaller $a$ it
does not allow to fix uniquely the dimensionality of
the chiral defects.

Moreover, the percolating strings shed light on the properties of the
 {\it stochastic defects}. Indeed, the strings percolate through the vacuum and any percolation is stochastic. 
Thus, strings are a source of stochastic fields which
 can be responsible, therefore, for the localization of the test scalar
 particles. 
 
 Most remarkable, the strings bring in a mixed scale, $\sqrt{\Lambda_{QCD}a^{-1}}$.
 In particular, the gluon condensate associated with
 the string is of order,
 \begin{equation}
 \langle G^{2}\rangle_{strings}~\sim~\Lambda_{QCD}^{2}a^{-2}~,
 \end{equation}
see Eq (\ref{singular}). This, mixed scale is manifested also in the properties of the scalar test particles
 in the adjoint representation, see Eq. (\ref{results}). Thus, localization of scalar particles might be related 
to the 2d defects.
Identifying the strings with the stochastic component results also in a successful prediction
of the confining string tension \cite{vz1}.

 \subsection{Fundamental 3d topological defects?}
 
 Thus, assumption on the fundamental role of the strings, or 2d defects seems to work 
very well in many cases. The only remaining challenge is
the dimensionality of chiral defects. Namely, measurements of the dimensionality of the topological 
defects mostly produce the value $\alpha_{5}=1$
 (see Eq (\ref{shrinking})) while for surfaces we would have $\alpha_{5}=2$. 
 Detailed exposition of evidence in favor of $\alpha_{5}=1$ 
 can be found in \cite{gubarev,reviews}.
 
 If indeed $\alpha_{5}=1$ then the surfaces cannot explain the data
 on the topological defects and   3d defects should
be considered as fundamental as well.

\section{Limit of classical solutions}

In the large $N_{c}$, supersymmetric version of the Yang-Mills theories
low-dimensional defects exist  as classical solution. In particular,
domain wall, or 3d defects carry chirality of the gluon field \cite{witten}.
One might speculate that quantum numbers of the defects,
such as dimensionality and chirality,  survive even  if the number of colors is not 
large. Then we would favor 3d topological defects to be fundamental,
that is irreducible to the 2d defects. 

\section{Conclusions}
Quite recently, it would sound heretic to talk about 
singular confining fields and non-trivial lattice-spacing dependences
of observables. Nowadays, the reality of the lattice measurements is such that
it is rather lack of lattice-spacing dependence of an observable that asks 
for explanation. 
Indeed, in the text we mentioned the following critical exponents:
which specify dependence on the lattice spacing, or resolution:
\begin{eqnarray}\label{striking}
\alpha_1\approx1,~\alpha_2\approx 3,~\alpha_4\approx 1,~\alpha_5\approx 2,
\alpha_6\approx 2,\\~\alpha (T=1)\approx 2,~\beta_1\approx1,~
\beta_2\approx1,~\beta(T=1)\approx 1~.\nonumber
\end{eqnarray}
Each of these indices is equal to zero in the standard picture of soft confining fields.

Thus,  
confining fields, as seen in measurements  with high 
resolution, are certainly singular. What is striking about the indices 
(\ref{striking}) is that they are --within error bars-- integer numbers.
Low-dimensional vacuum defects appear to 
be an adequate language to described the lattice data. 
The absence of anomalous fractal dimensions is remarkable and
we are encouraged to consider dimension of the defects as a quantum number.
Further characteristics are Abelianity and chirality of the non-Abelian 
fields living on the defects. 

Absence of anomalous fractal dimensions is natural in case of supersymmetric
theories with large number of colors. In this limit vacuum defects exist as classical
solutions and supersymmetry provides mechanism for cancellation
of quantum corrections. 
The lattice data (\ref{striking}) might indicate that basic geometric structures
survive in case $N_c=2,3$ as well.
Thus, it is not ruled out that the data indicate relevance of theory
of strings and/or other (fundamental) extended objects to physics of confinement
at short distances. If so, one could draw analogies to the developments in
particle physics which took place about 50 years ago.

\section*{Acknowledgments}

 We are thankful to O. Andreev, M.N. Chernodub, A, Di Giacomo,
L.B. Okun, M.I. Polikarpov,
T. Suzuki for useful discussions. 
and to K. Yamawaki for the invitation to the conference and hospitality.

%\bibiliographystyle{ws-procs9x6}

\end{document}